\UseRawInputEncoding
\documentclass[prd,aps,reprint,amsmath,amssymb,nofootinbib]{revtex4-1}
\pdfoutput=1
\usepackage{verbatim,graphics,graphicx,color,slashed,textcomp,bbm,mathdots}
\usepackage[normalem]{ulem} 
\usepackage{amsmath}   

\usepackage{natbib}    
\graphicspath{{figures/}}
\usepackage[utf8]{inputenc} 
\usepackage[T1]{fontenc} 
\usepackage{tabu} 
\usepackage{multirow} 
\usepackage[ruled]{algorithm2e} 
\PassOptionsToPackage{ruled,vlined,algo2e}{algorithm2e}
\usepackage{float} 
\usepackage[table]{xcolor}
\usepackage{makecell} 
\usepackage{booktabs} 
\usepackage{xcolor}
\usepackage[colorlinks=true,linkcolor=red,urlcolor=blue,citecolor=blue]{hyperref}
\begin{document}

\title{\text{\texttt{MatBYIB}}: A Matlab-based code for Bayesian inference of extreme mass-ratio inspiral binary with arbitrary eccentricity}

\author{Gen-Liang Li$^{a}$}
\author{Shu-Jie Zhao$^{b}$}
\author{Huai-Ke Guo$^{c}$}
\author{Jing-Yu Su$^{a,*}$}
\author{Zhen-Heng Lin$^{a,*}$}
\affiliation{\begin{footnotesize}
		${}^a$New Engineering Industry College, Putian University, Putian 3511000, China; \\
		${}^b$Key Laboratory of Particle Astrophysics, Institute of High Energy Physics, Chinese Academy of Sciences, Beijing 100049, China; \\
		${}^c$International Centre for Theoretical Physics Asia-Pacific, University of Chinese Academy of Sciences, Beijing 100190, China; 
\end{footnotesize}}

\begin{abstract}
Accurate parameter estimation is essential for gravitational wave data analysis. In extreme mass-ratio inspiral binary systems, orbital eccentricity is a critical parameter for parameter estimation. However, the current software for the parameter estimation of the gravitational wave often neglects the direct estimation of orbital eccentricity. To fill this gap, we have developed the \texttt{MatBYIB}, a MATLAB-based software package for the parameter estimation of the gravitational wave with arbitrary eccentricity. 
The \texttt{MatBYIB} employs the Analytical Kludge waveform as a computationally efficient signal generator and computes parameter uncertainties via the Fisher Information Matrix and the Markov Chain Monte Carlo. For Bayesian inference, we implement the Metropolis--Hastings algorithm to derive posterior distributions. To guarantee convergence, the Gelman--Rubin convergence criterion (the Potential Scale Reduction Factor $\hat{R}$) is used to determine sampling adequacy, with \texttt{MatBYIB} dynamically increasing the sample size until $\hat{R}<1.05$ for all parameters. Our results demonstrate strong agreement between predictions based on the Fisher Information Matrix and full MCMC sampling. This program is user-friendly and allows for the estimation of the gravitational wave parameters with arbitrary eccentricity on standard personal computers. \textit{Code availability:} The implementation is open-source at https://github.com/GenliangLi/\texttt{MatBYIB}.
\end{abstract}


\maketitle

\section{Introduction}\label{Introduction}
In gravitational wave (GW) detection and in many other areas, parameter estimation (PE) is one common part of the statistical analysis, with a goal of inferring the parameters of
the system generating the GW; using data recorded by ground-based detectors 
LIGO~\cite{Aasi2015, Abbott2021a}, Virgo~\cite{Acernese2015}, and KAGRA~\cite{Akutsu2020}; and future space-based GW detectors, including LISA~\cite{amaroseoane2017}, Taiji~\cite{Hu2017}, and Tianqin~\cite{Luo_2016}, etc
. For extreme mass-ratio inspiral (EMRI), the task of PE is to estimate, as precisely as possible, the masses, spins, sky locations, distance, etc., in order to
reveal the properties of the astrophysical population, test fundamental physics~\cite{krishnendu2021testing}, and probe possible new physics~\cite{Zhang:2021mks,Gondolo_1999,Li_2022}.

PE necessitates the construction of precise and rapid GW templates to facilitate the swift identification of signals within detector noise. For most LIGO GW sources, the orbital eccentricity is conventionally neglected under the quasi-circular approximation (e.g., the \texttt{Taylor} series~\cite{Damour2001, Blanchet2002, Blanchet2004}), and these computationally efficient models enable rapid GW PE. However, orbital eccentricity serves as a critical discriminator for probing the formation environments and mechanisms of binary systems~\cite{Nishizawa2016a,Nishizawa2016b,Breivik2016}. Its inclusion is essential for the accurate PE of GW sources~\cite{Sun2015}.
The development of precision templates incorporating eccentricity remains a significant challenge in contemporary GW astronomy.

The current prevalent methodologies for generating eccentric compact binary waveforms include the effective one-body (\texttt{EOB}, \texttt{SEOBNR})~\cite{Taracchini2014}, the frequency-domain phenomenological template series (\texttt{Phenom}~\cite{purrer2014frequency} and \texttt{IMRPhenomP}~\cite{hannam2014simple,Khan2016}), and the eccentric post-circular (\texttt{EPC})~\cite{Sun2015}. However, the EMRI with relatively large mass ratios ($10^4$ to $10^7$) requires a longer evolution time and the GW model with higher precision. Currently, the dominant models for EMRI are the Analytical Kludge (AK)~\cite{Barack2004}, Numerical Kludge (NK)~\cite{Babak2007}, and Augmented Analytical Kludge (AAK)~\cite{chua2017augmented} models. Although the AK waveform have many limitations, such as low accuracy~\cite{chua2017augmented}, its computational efficiency and capacity for term extension have enabled widespread application~\cite{Saltas2025}.  

PE generally requires the use of Bayes' theorem to obtain the posterior probability distribution of parameters.  Markov Chain Monte Carlo (MCMC) is the most commonly used method to obtain the posterior probability. There are several prominent, community-developed GW PE codes, including \texttt{LALInference}~\cite{LALSuite,swiglal}, \texttt{PyCBCInference}~\cite{Biwer2019}, 
and \texttt{Bilby}~\cite{RomeroShaw2020}. Common MCMC sampling tools include \texttt{emcee}~\cite{ForemanMackey2013}, \texttt{PyMC}~\cite{Salvatier2016}, \texttt{dynesty}~\cite{Speagle2020}, and \texttt{mcmcstat}~\cite{repec:boc:bocode:s457388}. These packages have been tested by multiple research groups and are widely recognized as standard tools in the field. However, open-access MCMC packages written in MATLAB for GW astronomy observations are not commonly seen.

Leveraging the advantages of MATLAB, including its real-time editing capabilities, rapid testing, interactive visualization environment, and highly concise code syntax, we have developed a MATLAB-based Bayesian inference toolkit for GW signals from EMRI with arbitrary eccentricities: \texttt{MatBYIB}. This tool employs the framework of the AK waveform to generate GW. \texttt{MatBYIB} derives posterior distributions of GW parameters through MCMC sampling and incorporates the convergence diagnostic method proposed by Gelman and Rubin~\cite{gelman1992inference}, which defines a Potential Scale Reduction Factor $\hat{R}$, and the MCMC sampling continues until $\hat{R} < 1.05$~\cite{gelman1992inference} for all parameters. 
 The entire program is implemented with multiple independent MCMC chains running in parallel to enhance the computational efficiency and robustness in the sampling process. This design ensures that the toolkit can handle complex and computationally intensive tasks efficiently, making it suitable for large-scale parameter estimation and predictions. In conclusion, \texttt{MatBYIB} provides a simple, efficient, and scalable code for PE and predictions for future GW detectors such as LISA and Taiji. 
 
\section{Theroy}\label{Theroy}
\subsection{Waveform Generation}\label{waveform generation}
A compact stellar-mass object $m_2$
  (typically a stellar black hole (BH) or a neutron star (NS)) orbits around a central supermassive BH of mass $m_1$, and this system emits GW in the millihertz frequency band, which is called EMRI. The AK waveform is currently one of the most widely used GW templates for EMRI. Although AK waveforms exhibiting errors are approximately an order of magnitude larger than those of the AAK waveforms~\cite{chua2017augmented},
they are convenient to use for simple parameter error estimation because of their lower computational complexity, and they are good at simulating the motion of particles far from the BH using post-Newtonian approximation.

The AK waveform is mainly composed of two parts. The first part describes the orbital dynamics of the small object (Effective Single-Body Approximation assumes that large central BH is stationary~\cite{Buonanno_1999}
), and the second part is waveform production via the Peters--Mathews waveform equation under the quadrupole approximation~\cite{Peters1963}.

The evolution of GW orbital parameters with time can be expressed as 
 \cite{Barack2004}
 \vspace{-12pt}
 \begin{widetext}
\begin{align}\label{eq:alpha_t}
\frac{d\Phi}{dt} =  &2\pi f_{orb}, 
\nonumber\\
\frac{df_{orb}}{dt} =& (\frac{96}{10\pi}) (\frac{c^6\mu}{G^2M^3} ) (\frac{2\pi G M f_{orb}}{c^3})^{11/3} (1 - e^2)^{-9/2} [(1 - e^2) (1 + \frac{73}{24} e^2 + \frac{37}{96} e^4)\nonumber\\
&+ (\frac{2\pi G M f_{orb}}{c^3})^{2/3} (\frac{1273}{336} - \frac{2561}{224} e^2 - \frac{3885}{128} e^4 - \frac{13147}{5376} e^6) \nonumber\\
& - (\frac{2\pi G M f_{orb}}{c^3}) (\frac{c}{G} \frac{S}{M^2}) \cos\lambda
(1 - e^2)^{-1/2} (\frac{73}{12} + \frac{1211}{24} e^2 + \frac{3143}{96} e^4 + \frac{65}{64} e^6)],
\nonumber\\
\frac{de}{dt}= &-\frac{e}{15} (\frac{c^3\mu}{GM^2} ) (\frac{2\pi G M f_{orb}}{c^3})^{8/3} (1 - e^2)^{-7/2} [(1 - e^2)(304 + 121 e^2)(1 + 12 (\frac{2\pi G M f_{orb}}{c^3})^{2/3}) \nonumber\\
&- \frac{1}{56} (\frac{2\pi G M f_{orb}}{c^3})^{2/3} (8 \cdot 16705 + 12 \cdot 9082 e^2 - 25211 e^4)] + e \frac{c^3\mu}{GM^2} (S) \cos\lambda \nonumber\\
&(\frac{2\pi G M f_{orb}}{c^3})^{11/3} (1 - e^2)^{-4} (\frac{1364}{5} 
+ \frac{5032}{15} e^2 + \frac{263}{10} e^4),
\nonumber\\
\frac{d\tilde{\gamma}}{dt} =  &6\pi f_{orb} (\frac{2\pi G M f_{orb}}{c^3})^{2/3} (1 - e^2)^{-1} (1 + \frac{1}{4} (\frac{2\pi G M f_{orb}}{c^3})^{2/3} (1 - e^2)^{-1} (26 - 15 e^2))\nonumber\\
&- 12\pi f_{orb} (\frac{c}{G}\frac{S}{M^2}) \cos\lambda(\frac{2\pi G M f_{orb}}{c^3}) (1 - e^2)^{-3/2},
\nonumber\\
\frac{d\alpha}{dt} = &4\pi f_{orb} (\frac{2\pi G M f_{orb}}{c^3}) (\frac{c}{G} \frac{S}{M^2}) (1 - e^2)^{-3/2}.\\
\nonumber
\end{align}
 \end{widetext}
 \vspace{-12pt}
where $M$, $\Phi$, $f_\textrm{orb}$, and $e$ are the total mass, orbital phase, frequency, and eccentricity, respectively. $\tilde{\gamma},\alpha$ are the two precession angles. We can also include higher-order PN terms if necessary.

We establish a Cartesian coordinate system in the ecliptic plane, with the axes denoted as \( x \), \( y \), and \( z \). The spin angular momentum of the BH is represented by the vector \( \hat{S} \), where \( S \) signifies the magnitude of the angular momentum. Meanwhile, the orbital angular momentum is represented by the vector \( \hat{L}(t) \), whose orientation is determined by the inclination angle \( \lambda \)---the angle between the vectors \( \hat{L} \) and \( \hat{S} \)---and the azimuthal angle \( \alpha(t) \). The projection of the orbital angular momentum in the direction of wave propagation can be represented as
\begin{align}
    \hat{L} \cdot \hat{n}= & \hat{S} \cdot \hat{n} \cos \lambda+\frac{\cos \theta_{S}-\hat{S} \cdot \hat{n} \cos \theta_{K}}{\sin \theta_{K}} \sin \lambda \cos \alpha \nonumber \\ &+\frac{(\hat{S} \times {z}) \cdot \hat{n}}{\sin \theta_{K}} \sin \lambda \sin \alpha,
\end{align}
where $\theta_K$ and $\phi_K$ are the sky orientation angles of the spin angular momentum vector.  If the center BH is the Schwarzschild BH, the angle between $\tilde{L}$ and $\tilde{S}$ is $\lambda = 0$. For the $n$-th harmonic wave, \( h_n^+ \) can be written as
\begin{align}\label{eq:h_n}
 h_{n}^{+}=&-\frac{1}{D}\{[ 1+(\hat{L}\cdot \hat{n})^2 ] \left[ a_n\cos\mathrm{(}2\xi )-b_n\sin\mathrm{(}2\xi ) \right]\nonumber\\
 &+[ 1-(\hat{L}\cdot \hat{n})^2]c_n\}, \\
 h_{n}^{\times}=&\frac{2}{D}(\hat{L}\cdot \hat{n})[ b_n\cos(2\xi)+a_n\sin(2\xi)],   
\end{align}
in which $\xi$ is an azimuthal angle used to measure the direction of the pericenter relative to the $x$ axis~\cite{Barack2004}, and $\hat{x}$ is defined as
$\hat{x} \equiv \frac{-\hat{n} + \hat{L} (\hat{L} \cdot \hat{n})}{\sqrt{1 - (\hat{L} \cdot \hat{n})^2}}
$. $D$ is the luminosity distance, and $a_n$, $b_n$, and $c_n$  are the superpositions of Bessel functions related to the eccentricity~\cite{Barack2004}. As for the circular orbital case, $a_n,b_n$ is zero. Therefore, we have obtained the GW source signal by the superposition of harmonic waves.

\subsection{Response Function}\label{detector}
To analyze the GW source, it is necessary to transform the GW source into the detector's reference frame. 
The detection of GW is based on the minor relative changes $\delta L$ in the lengths of two arms of Michelson interferometers ~\cite{Cutler_1998,Tinto2002,Barack2004}; we refer to the two-arm detector formed by arms $1$ and $2$ as the ``detector I'' and the ``detector II'', and the GWs are expressed as \( h_{I}(t) =\left[\delta L_{1}(t)-\delta L_{2}(t)\right] / L \), and $h_{II}(t)=3^{-1 / 2}\left[\delta L_{1}(t)+\delta L_{2}(t)-2 \delta L_{3}(t)\right] / L$
, where \( L \) denotes the average arm length of the detector. Given that the space-based GW detectors comprise three arms forming an equilateral triangular, they essentially function as a pair of Michelson interferometers, and the additional factor of \( \sqrt{3}/2 \) arises because the angle between the arms of the space-based detector is $\pi/3$ rather than $\pi/2$.  
The representation of GWs in the  space-based detector frame has been formulated as~\cite{chua2017augmented}
\begin{equation} 
\label{eq:ht}
h_{I,II}(t)=\frac{\sqrt{3}}{2}\left[F_{I,II}^{+}(t) h_{I,II}^{+}(t)+F_{I,II}^{\times}(t) h_{I,II}^{\times}(t)\right],
\end{equation} 
where $F^+$ and $F^{\times}$ are the response functions. 
\begin{align}
\label{eq:F_+_LISA}
F_{I,II}^{+}(\theta, \phi, \psi)=&{[\frac{1}{2}\left(1+\cos ^{2} \theta\right) \cos 2 \phi \sin 2 \psi} \\ \nonumber
 & \substack{- \\ +}\cos \theta \sin 2 \phi \cos 2 \psi],
 \end{align}
\begin{align}
\label{eq:F_x_LISA}
F_{I,II}^{\times}(\theta, \phi, \psi)=&{[\frac{1}{2}\left(1+\cos ^{2} \theta\right) \cos 2 \phi \sin 2 \psi} \\ \nonumber
&\substack{+ \\ -} \cos \theta \sin 2 \phi \cos 2 \psi],
  \end{align}
where $(\theta,\phi)$  are the sky location and  $\psi$  is  the polarization angle of the GW source~\cite{Barack2004}. For space-based detectors, the sky angles $ \theta,\phi,\psi$ in the detector frame change over time due to the continuous rotation of the detectors~\cite{chua2017augmented}.
By performing the Fourier transform on Equation~\eqref{eq:ht}, we can obtain the GW in the frequency domain $\tilde{h}(f)$.

\subsection{Fisher Information Matrix}\label{sec:FM}
The Fisher Information Matrix (FIM) provides a computationally efficient framework for the rapid precision estimation of GW parameters by quantifying the local curvature of the likelihood surface in the parameter space. As shown in Table~\ref{tab:parameters}, the GW from two inspiraling bodies can be described by a set of parameters
 $\bm{\theta} = (\theta^1, \ldots, \theta^k)$; we consider this set of waveforms as a multidimensional surface embedded in the vector space of all possible measured signals~\cite{Cutler_1994}. The maximum likelihood estimator is indeed the value of \( \bm{\theta} \) that provides the highest signal-to-noise ratio (SNR) in the matched filtering~\cite{finn2000gravitational}. To compute the likelihood function, and hence the posterior probability, we assume for simplicity that the noise $n(t)$ is stationary and Gaussian. Given the detected signal ${s}(t) = {h}(t;\bm{\theta}) + {n}(t)$, where ${h}(t;\bm{\theta})$ is the GW template depending on parameters and ${n}(t)$ is the detector noise, the likelihood $\Lambda({s}\mid \bm{\theta} ) $ is
\begin{equation} \label{eq:likelihood}
\Lambda({s}\mid \bm{\theta}  ) = K \exp\left[ -\frac{1}{2} \left( {s} - {h}(\bm{\theta})\mid  {s} -{h}(\bm{\theta}) \right) \right],
\end{equation}
the $K$ is the normalizing constant, the inner product above is defined as $(n\mid n)=\mathrm{Re}\int_{-\infty}^{\infty}{d}f\frac{\tilde{n}^*(f)\tilde{n}(f)}{(1/2)S_n(f)}$, and $S_n$ is the sensitive curve of the detector. 

In the Bayesian approach
\begin{equation} \label{eq:posterier}
P(\bm{\theta} \mid {s}) =  \frac{P^{(0)}(\bm{\theta}) \Lambda( {s} \mid \bm{\theta})}{P(s)},
\end{equation}
where $P(\bm{\theta} \mid {s}) $ is the posterior distribution, $P^{(0)}(\bm{\theta})$ is the prior distribution, and ${P(s)}$ is the evidence. We can expand the $(s-h\mid s-h)$ around the maximum likelihood estimator $\bm{\theta}_\mathrm{ML}$ and the linear term of the expansion vanishes. We can obtain~\cite{finn2000gravitational}
\begin{equation}
 P((\bm{\theta} \mid s) = K \exp \left\{ -\frac{1}{2}\Gamma _{ij}\Delta \theta ^i\Delta \theta ^j \right\},
 \end{equation} 
where ${\Gamma_{i j}=\left(\partial_{i} \partial_{j} h \mid h-s\right)+\left(\partial_{i} h \mid \partial_{j} h\right)}$. In the high SNR (e.g., SNR > 10~\cite{Cutler_1994}) regime, $(h-s) \ll h$ and 
\begin{equation}
\label{eq:gamma_vuale}
 \Gamma _{ij}=\left( \partial _ih\mid \partial _jh \right),
 \end{equation} 	
the $\Gamma$ is FIM, the covariance matrix is given by
$\left. \langle \Delta \theta ^i\Delta \theta ^j \right. \rangle =\left( \Gamma ^{-1} \right) ^{ij}$, and  the expected value of parameter error $\Delta \theta ^i$ is given by
	$\Delta \theta ^i=\sqrt{\left. \langle \Delta \theta ^i\Delta \theta ^i \right. \rangle}=\sqrt{\left( \Gamma ^{-1} \right) ^{ii}}$.
	
\begin{table*}[htbp]
  \centering
  \caption{Summary of physical parameters and their meaning. The angles $\left(\theta_{S}, \phi_{S}\right)$ and $\left(\theta_{K}, \phi_{K}\right)$ are associated with a spherical coordinate system attached to the ecliptic. $\hat{L}$ and $\hat{S}$ are unit vectors in the directions of the orbital angular momentum and the MBH's spin, respectively.}  %
  \label{tab:parameters}  %
  \resizebox{\linewidth}{!}{
  \begin{tabular}{ccccc}
    \toprule
   line number & Values  & Units & Parameters & physical quantity \\  
    \midrule
   $1$ & $1.e+6$ &  $M_{\odot}$ & $m_1$& the mass of central BH\\  
   $2$ & $10$ &  $M_{\odot}$ & $m_2$ &the mass of rotating object\\
   $3$ & $0.3$ &  $--$ & $e_\mathrm{LSO}$ & where $e_\mathrm{LSO}$ is the last stable circular orbital eccentricity\\
     $4$ & $0.01$ &  $ --$ & $S/M^2$ &magnitude of spin angular momentum of
MBH\\
   $5$ & $0.01$ &  $--$ & $z$ &red shift\\
   $6$ & $60$ &  $^\circ$ & $\lambda$& $cos\lambda= {\hat{L} \cdot \hat{S}}$\\
   $7$ & $0$ &  $--$ & $\phi_\mathrm{LSO}$ &where $\phi_\mathrm{LSO}$ is the last stable circular orbital mean anomaly\\
   $8$ & $60$ &  $^\circ$ & $\gamma_\mathrm{LSO}$ &where
$\gamma_\mathrm{LSO}$ is the angle (in orbital plane) between $\hat{L} \times \hat{S}$ and pericenter\\
   $9$ & $60$ &  $^\circ$ & $\alpha_\mathrm{LSO}$ &where $\alpha_\mathrm{LSO}$ is the azimuthal direction of $\hat{L}$ in the orbital plane\\
   $10$ & $60$ &  $^\circ$ & $\theta_S$ &the source direction’s polar angle\\
   $11$ & $60$ &  $^\circ$ & $\phi_S$ &azimuthal direction to source\\
   $12$ & $60$ &  $^\circ$ & $\theta_k$ &the polar angle of MBH’s spin\\
   $13$ & $60$ &  $^\circ$ & $\phi_k$ &azimuthal direction of MBH’s spin\\
   $14$ & $0$ &  $^\circ$ & $\phi_0$ &The mass of rotating object\\
   $15$ & $3.14e+6$ &$s$ &  $t_{c}$  & $t_c$ is time where orbit is last stable circular orbit\\
\bottomrule
\end{tabular}}
\end{table*}

\subsection{Markov Chain Monte Carlo}
\label{MCMC}
The MCMC method is a technique for sampling from complex probability distributions by constructing a Markov chain. The core thought is to design a Markov chain whose samples converge to the target distribution. Various sampling methods have been developed within the MCMC framework, including Metropolis--Hastings (M-H)~\cite{Metropolis1953, Hastings1970}, No-U-Turn sampling~\cite{Homan2014}, Gibbs sampling~\cite{Muller1991, Gilks1992, Ritter1992}, Blocked M-H~\cite{Andrieu2003}, Parallel Tempering~\cite{Swendsen1986, Hukushima1996}, and Reversible Jump. 

In our work, we employ the M-H sampling method. 
The process begins by selecting a random position in the parameter space. For the \((i+1)\)-th step, the position of \( \bm{\theta}_{i+1}\) is determined only based on the previous point, \( \bm{\theta}_i\). Typically, this is achieved by constructing a transition matrix, \( Q(\bm{\theta_{i+1}} \mid \bm{\theta}_i)  = q(\bm{\theta_{i+1}} \mid \bm{\theta}_i) c(\bm{\theta_{i+1}} \mid \bm{\theta})\), where $q(\bm{\theta_{i+1}} \mid \bm{\theta}_i)$ denotes the proposal distribution and $c(\bm{\theta_{i+1}} \mid \bm{\theta})$ represents the acceptance probability. The proposal distribution may take various forms~\cite{roberts2004general}, including uniform distributions, an independence sampler, or gradient-based distributions derived from the target distribution~\cite{roberts1996exponential,duane1987hybrid}, with different choices significantly affecting the sampling efficiency. In M-H sampling, Gaussian proposals are most commonly adopted due to the local Gaussian approximation property of posterior distributions, particularly near their modes, which ensures that the sampling density remains proportional to the posterior probability distribution~\cite{tiertney1994markov}. Here we assume that the proposal distribution satisfies the Gaussian distribution~\cite{Andrieu_2008}
\begin{equation}
q(\bm{\theta}_{i+1}\mid \bm{\theta}_i)=\frac{1}{\sqrt{2\pi \sigma^{2}}} e^{-\frac{(\bm{\theta}_{i+1}-\bm{\theta}_i)^{2}}{2 \sigma^{2}}},
\end{equation}
where ${\sigma}$ is a covariance matrix, which we wish to be as small as possible for efficiency purposes. A candidate point is randomly drawn by the transition matrix \( q(\bm{\theta}_{i+1} \mid \bm{\theta}_i)\). Then, we need to calculate the acceptance probability $c({i+1},i)$,
\begin{align}\label{eq:M-H}
c(i+1,i) &= \min\{ 1, \frac{P(\bm{\theta}_{i+1})q(\bm{\theta}_{i+1} |\bm{\theta}_i) }{P(\bm{\theta}_i)q(\bm{\theta}_i | \bm{\theta}_{i+1})} 
\}.
\end{align} 
where the $P(\bm{\theta}_{i+1})$ is the poster distribution obtained from Equation~\eqref{eq:posterier}. To make the Markov chain converge more quickly, a random value $u \sim \text{uniform}(0,1)$ is typically generated~\cite{Hastings1970}. If $u < c({i+1},i)$, then retain $P(\theta_{i+1})$, set $P(\theta_i) = P(\theta_{i+1})$, and update. If $u > c({i+1},i)$, then retain $P(\theta_i)$ and repeat the process. The pseudocode of the M-H algorithm is given in Algorithm~\ref{Algorithm1}.

\begin{algorithm}[H]
\label{Algorithm1}
\caption{Pseudo-code of Metropolis-Hastings}
\KwIn{Initialize the Markov chain state $\bm{\theta}$, the number of iterations $N$}
\KwOut{The list of samples}
\BlankLine
\While{$i \leq N$}{
    Calculate the target distribution $P(\bm{\theta}_i)$ at the current state using Eq.~\eqref{eq:posterier}\;
    Generate a proposal $\bm{\theta}_{i+1}$ by the transition kernel $q(\bm{\theta}_{i+1} \mid \bm{\theta}_{i})$\;
    Compute the acceptance probability $c(i+1,i)$ using Eq.~\eqref{eq:M-H}\;
    Generate a uniform random number $u \sim \text{uniform}(0, 1)$\;
    \If{$\log(u) < c(i+1,i)$}{
        Accept the proposal: $\bm{\theta}_i = \bm{\theta}_{i+1}$\;
    }
    $i \leftarrow i + 1$\;
}
\end{algorithm}

\vspace{-12pt}
\subsection{Convergence Diagnostics}
We use the Gelman--Rubin method~\cite{gelman1992inference} to optimize the convergence criteria of MCMC. After discarding the burn-in period of the Markov chains, each chain is split into two parts; we assume that there are \( 2k \) Markov chains, each containing \( n \) iterations, and the value of the \( i \)-th iteration of the \( j \)-th chain for the target parameter \( \theta \) is denoted as \( \theta_{i,j} \). The variance between chains \( B \) reflects the differences between the means of different chains,
\begin{equation}\label{eq:between-chains}
B = \frac{n}{2k-1} \sum_{j=1}^{2k} (\bar{\theta}_j - \bar{\theta})^2,
\end{equation} 
where \( \bar{\theta}_j \) is the sample mean of the \( j \)-th chain, and \( \bar{\theta} \) is the overall mean.  The within-chain variance (\( W \)) characterizes the variability within a single chain,
\begin{equation} \label{eq:within-chains}
W = \frac{1}{2k} \sum_{j=1}^{2k} \left( \frac{1}{n-1} \sum_{i=1}^{n} (\theta_{i,j} - \bar{\theta}_j)^2 \right).
\end{equation}

By combining \( B \) and \( W \) through weighted integration, we obtain the variance estimate for the target parameter \( \theta \),
\begin{equation}\label{eq:total-variance}
\text{var}(\theta) = \left( \frac{n-1}{n} \right) W + \left( \frac{1}{n} \right) B,
\end{equation} 
under stationarity or as \( n \to \infty \), \( \text{var}(\theta) \) is an unbiased estimate and approximately equal to \textit{W}. The Potential Scale Reduction Factors are defined as
\begin{equation}\label{eq:Potential Scale Reduction Factors}
\hat{R}  = \sqrt{\frac{\text{var}(\theta)}{W}},
\end{equation} 
when \( \hat{R} \approx 1 \), the chain has reached stationarity; when \( \hat{R}  > 1.05\)~\cite{gelman1992inference}, the chain has not converged, and the number of iterations \( n \) needs to be increased. We set the program to automatically add the value of \( n \) and continue sampling.
\section{Software Architecture}

\begin{figure*}[htbp]
    \centering
        \centering
        \includegraphics[width=\linewidth]{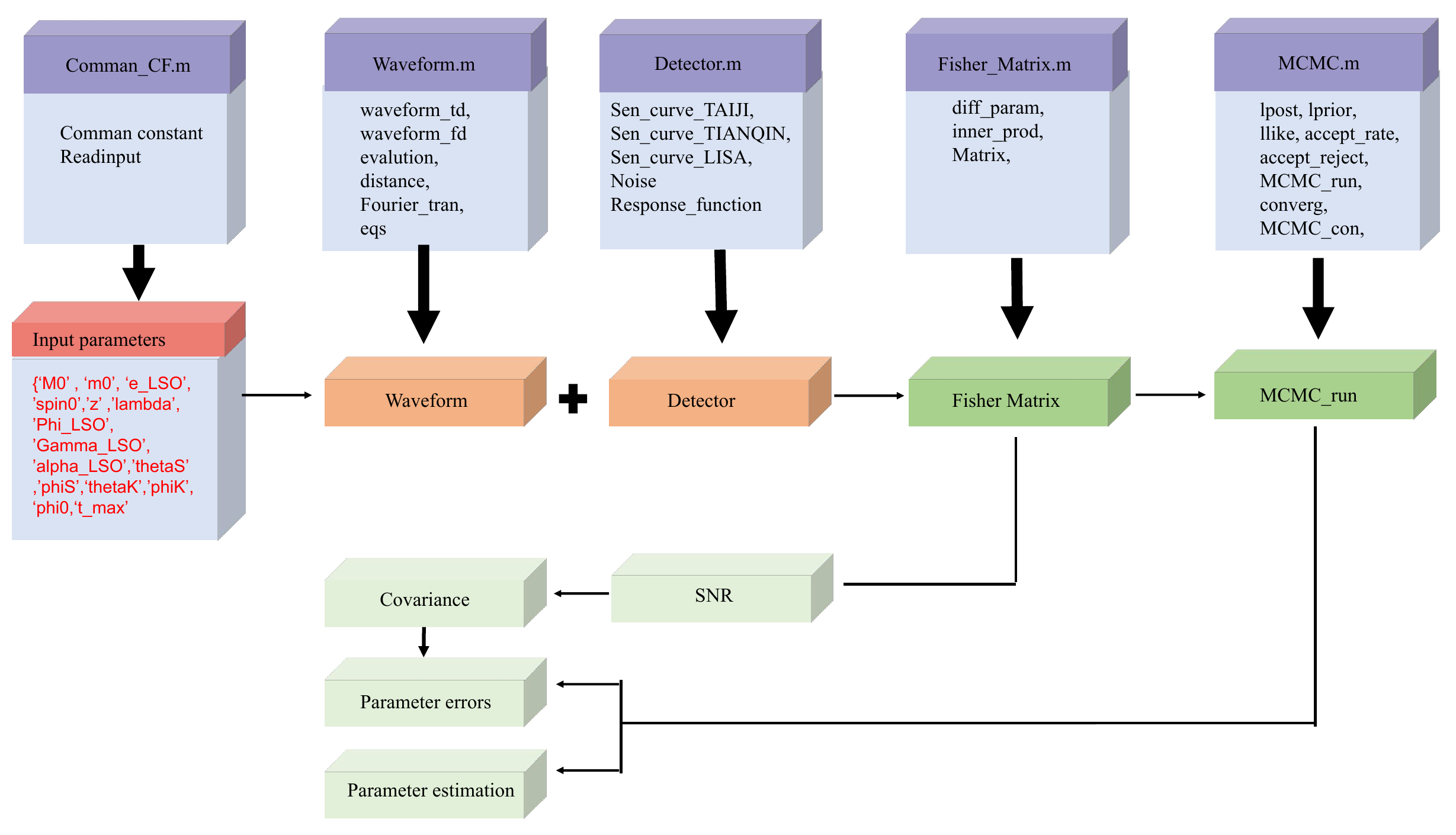}
        \caption{Schematic Diagram of the Software Architecture }
        \label{fig: Schematic Diagram}
\end{figure*}

As shown in Figure~\ref{fig: Schematic Diagram}, the software is divided into five major modules, namely, the common constants module (\textbf{Common\_CF.m}), 
the waveform generation module (\textbf{Waveform.m}), the detector module (\textbf{Detector.m}), the FIM module (\textbf{Fisher\_Matrix.m} and \textbf{MCMC.m}). The common constants module (\textbf{Common\_CF.m}) primarily encompasses common constants (\texttt{Common\_constants.m}) and the input parameter function (\texttt{Readinput( )}). The waveform generation module (\textbf{Waveform.m}) provides GW waveform templates, and the detector module offers functions related to the detectors, such as the detector response functions, noise function, and sensitivity curves. Using the FIM module, we can initially calculate the SNR of GW and the covariance of parameters, ultimately yielding parameter errors. Furthermore, the posterior distribution of the parameters can be obtained through the \texttt{MCMC\_run( )} function within the MCMC module (\textbf{MCMC.m}),

\noindent \textbf{Common\_CF.m}:
This module defines the values of physical constants and the function to read input files (\texttt{Readinput( )}). Users can fill in the corresponding parameters as needed for their models.\\

\noindent\textbf{Waveform.m}: This module defines a set of orbital evolution functions (see Equation~\eqref{eq:alpha_t}) and generates waveforms according to Equation~\eqref{eq:h_n}. The main functions are as follows:\\

\noindent\texttt{eqs( )}: Defines a set of ordinary differential equations (ODEs) (Equation~\eqref{eq:alpha_t}) and returns the values of the orbital parameters as they evolve. Note that there may be slight differences in the ODE solvers due to the different versions of MATLAB.\\

\noindent\texttt{evolution( )}: Calculates the solutions of the kludge ODE equations defined in \texttt{eqs( )}. To solve the  ODE system, we utilize the \texttt{ODE45} function in MATLAB. This choice facilitates the switch between different native solvers available in the library.\\

\noindent\texttt{get\_Aplus\_A\_cross( )}: This calls \texttt{evolution( )} and calculates the intensity of the different-order harmonic GW in the source coordinate system according to the Equation~\eqref{eq:h_n}.\\

\noindent\texttt{waveform\_td( )}: This calls \texttt{eqs( )}, \texttt{evolution( )}, and \texttt{get\_Aplus\_A\_cross( )}, to compute the time-domain GW including the detector response function from the detector module and then performs its Fourier transform via the \texttt{Fourier\_tran( )} function in Common\_CF.m. This process is integrated into the functions \texttt{Fisher\_Matrix( )} and \mbox{\texttt{MCMC\_run( )}} detailed below.\\

\noindent\textbf{Detector.m}: This module defines the sensitivity curves of various GW detectors, including LISA~\cite{Robson_2018ifk}, Taiji~\cite{Hu2017}, and Tianqin~\cite{Luo_2016}, as well as their response functions. Users can modify the detector sensitivity curve directly in the main.m file. For example, they can call the \texttt{Sen\_curve\_LISA( )} function from the Detector.m module to obtain the LISA \mbox{sensitivity curve.}\\

\noindent\texttt{get\_noise( )}: This defines the detector noise function, and the noise satisfies the \mbox{Gaussian distribution.}\\

\noindent\textbf{Fisher\_Matrix.m}:
This module defines functions related to the calculation of the \mbox{FIM, including:}\\

\noindent\texttt{Diff\_param( )}: This function defines the partial derivatives of the GW signal with respect to different parameters. It generates the frequency domain function by calling the \noindent\texttt{waveform\_fd( )} and uses the central difference to obtain the partial derivatives of the GW signal with respect to different parameters.\\

\noindent\texttt{Matrix( )}: This calls \texttt{diff\_param( )} and \texttt{inv( )} functions, with the latter being MATLAB's built-in function for calculating the inverse of a matrix.\\

\noindent\textbf{MCMC.m}: This module consists of functions related to MCMC. The functions \mbox{mainly include:}\\

\noindent\texttt{lprior( )}, \texttt{lpost( )}: These functions define the prior distribution and the posterior probability, respectively. They can be manually set to specific ranges.\\

\noindent\texttt{llike( )}: This function computes the likelihood according to Equation~\eqref{eq:likelihood}. It is used in \texttt{MCMC\_run( )} to compute the likelihood at each step.\\

\noindent\texttt{accept\_reject( )}: This function corresponds to Equation~\eqref{eq:M-H}.\\

\noindent\texttt{MCMC\_run( )}: This calls \texttt{waveform\_fd( )} to generate the GW signal, calculates the posterior distribution at each step according to Equation~\eqref{eq:likelihood}, and calls \texttt{accept\_reject( )} to determine whether to retain the current particle's posterior distribution value, Equation~\eqref{eq:M-H}, and then continues to iterate.\\

\noindent\texttt{converg( )}: The function uses Equations~\eqref{eq:between-chains}--\eqref{eq:total-variance} to assess the convergence of sampling. If $\hat{R}> 1.05$, the chain is considered non-convergent, and we will increase the sampling points. Typically, in our program, the number of points is increased to 1.25 times the original total number, after which \texttt{MCMC\_contin( )} is invoked to resume sampling.\\
 
\noindent\texttt{MCMC\_contin( )}: Its usage is identical to that of \texttt{MCMC\_run( )}. 

\section{Test and Numerical Examples}
GW parameters can be categorized into intrinsic and extrinsic parameters. Extrinsic parameters describe the observer's reference frame, including the coalescence time ($t_{max}$), 
sky position angles ($\theta_{\rm S}$, $\phi_{\rm S}$), luminosity distance ($D$), polarization angles ($\psi_{\rm K}$, $\phi_{\rm K}$), and phase at the last stable orbit ($\phi_{LSO}$). In contrast, intrinsic parameters characterize the source physics independently of the observer's orientation, such as the component masses ($m_1$, $m_2$), dimensionless spin ($\frac{S}{M^2}$), 
eccentricity at the last stable orbit ($e_{\mathrm{LSO}}$), cosine of inclination ($\cos\lambda$), and dimensionless pericenter advance ($\tilde{\gamma}_{LSO}$). Extrinsic parameters are computationally less expensive to sample compared to intrinsic parameters due to their weaker influence on waveform morphology. Our framework can compute posteriors for all parameters, and we restrict the analysis to four intrinsic parameters 
($m_1$, $m_2$, $\frac{S}{M^2}$, and $e_{\mathrm{LSO}}$) in the following test cases to reduce the computational cost.

The implementation details, input parameter specifications, and considerations concerning numerical precision are addressed in the accompanying source code and program documentation. Therefore, this section presents only selected numerical examples for demonstration purposes.

\begin{figure*}[htp]
    \centering    
    \includegraphics[width=0.95\textwidth]{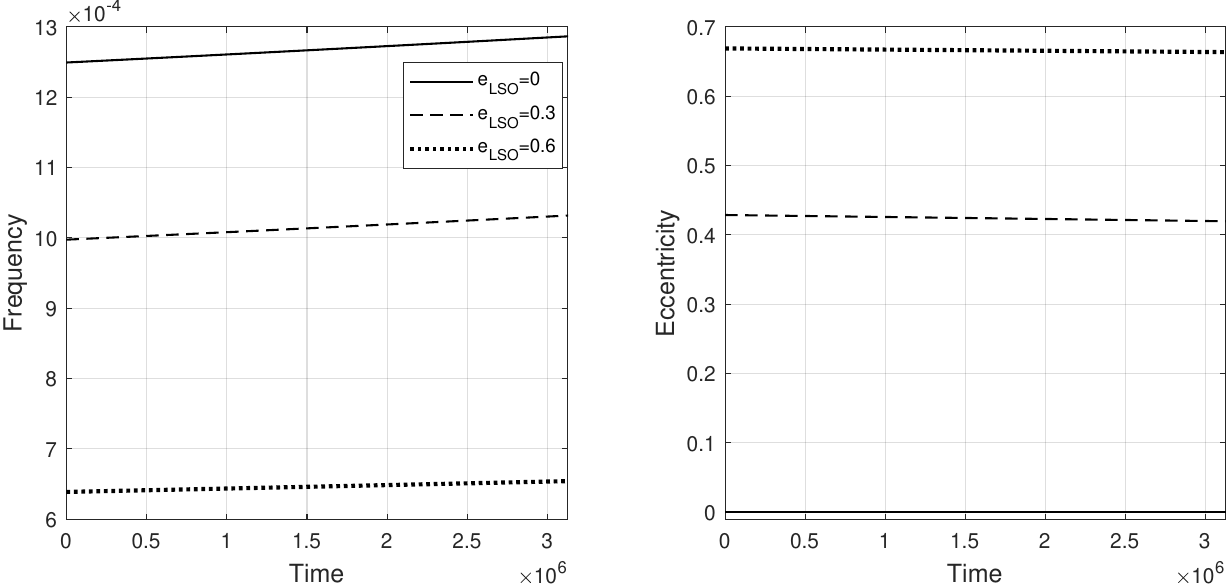}
    \caption{The evolution of orbital frequency (Left) and eccentricity (Right) with time. The system parameters are as follows: the mass of the central BH is \( M = 10^6 M_{\odot} \), the mass of the orbiting object is \( 10 M_{\odot} \), the spin parameter of the central BH is \( S/M^2 = 0.4 \), the eccentricity at the last stable orbit \( e_\mathrm{LSO} = 0.0,0.3,0.6 \), and the frequency at the last stable orbit is calculated by \({f_\mathrm{LSO}}=c^{3} /(2 \pi G M)\left(\left(1-e_{\mathrm{LSO}}^{2}\right) /\left(6+2 e_{\mathrm{LSO}}\right)\right)^{3 / 2}\). The evolution time is \( t_c = 3.14 \times 10^6 \) seconds.}
    \label{Fig:evolution}
\end{figure*}

\begin{figure*}[htbp]
\includegraphics[width=0.325\textwidth]{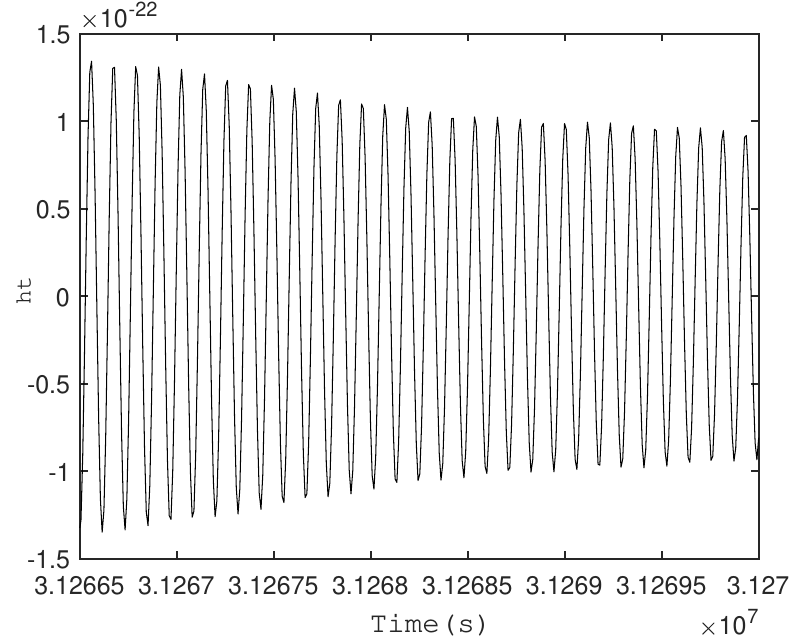}   
\includegraphics[width=0.325\textwidth]{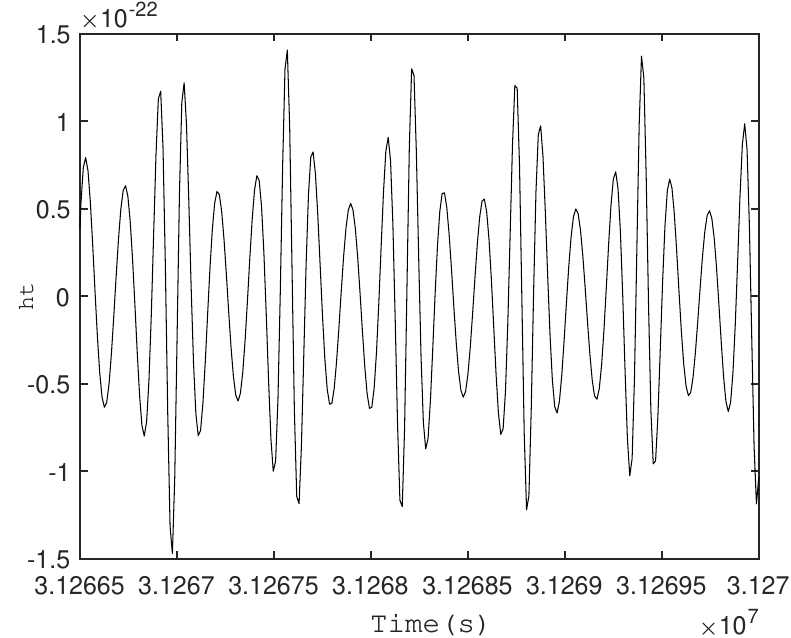}   
\includegraphics[width=0.325\textwidth ]{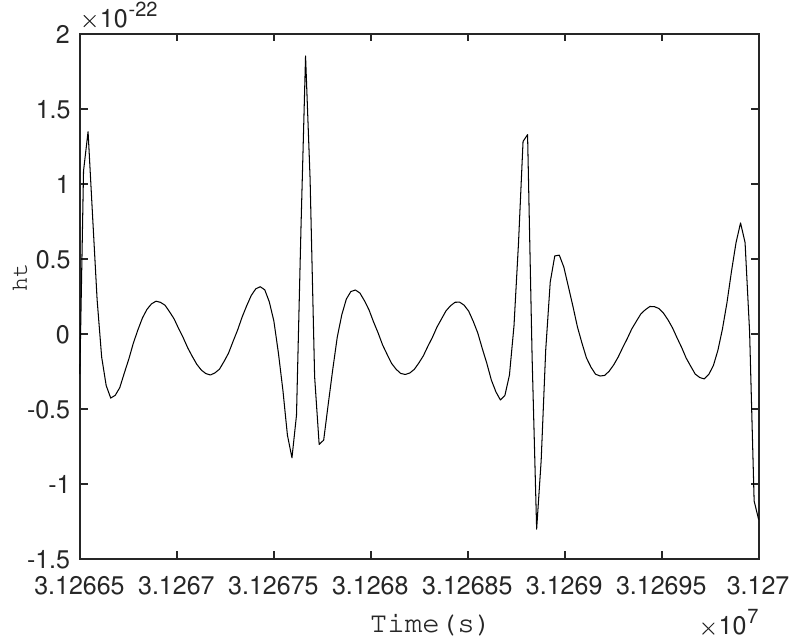}
    \caption{The time-domain gravitational waveforms \textit{h}(\textit{t}) are shown for three distinct time intervals: ${0 \sim 3000}$ (left panel), ${1.109\times10^6 \sim 1.204\times10^6}$ (middle panel) and ${3.137\times10^6 \sim 3.140\times10^6}$ (right panel). The system parameters are configured as follows: CO's mass: ${m_2=10 M_{\odot}}$; MBH's mass: ${M=10^{6} M_{\odot}}$; MBH's spin magnitude: ${S=0.01 M^{2}}$; Angle between MBH's spin and orbital angular momentum: ${\lambda=60{^\circ}}$; We set $\phi_\mathrm{LSO} = \gamma_\mathrm{LSO} = \alpha_\mathrm{LSO} = 0 $;
    Sky angle $\theta_S = \phi_S = \theta_k =\phi_k = 60 ^\circ$.}
    \label{fig:ht}
\end{figure*}

\begin{figure*}[hbtp]
\includegraphics[width=0.485\textwidth]{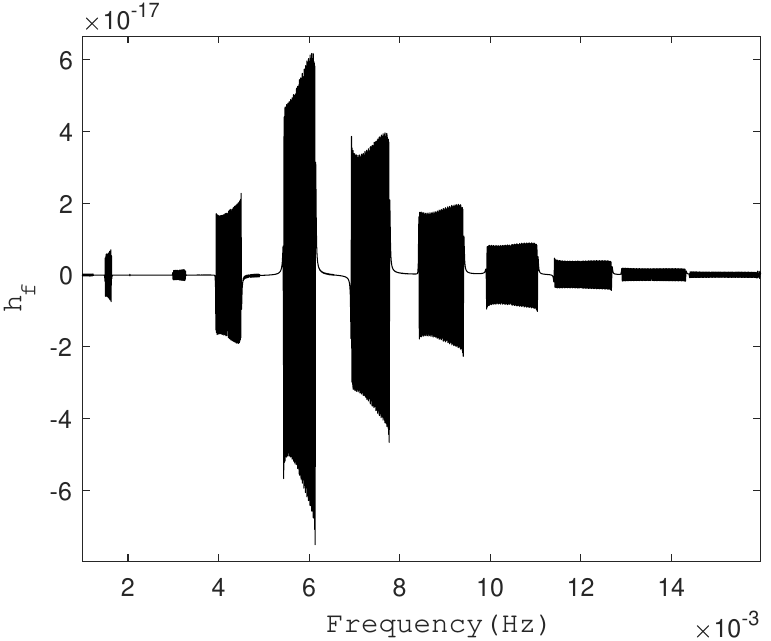}   
\includegraphics[width=0.485\textwidth]{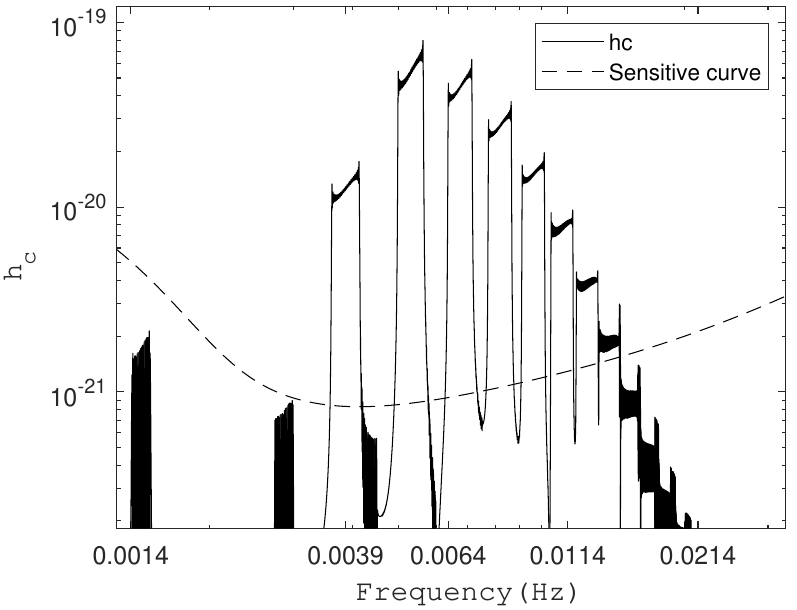} 
\caption{GW signal on frequency domain. The frequency-domain GW waveforms by FFT on time-domain GW in Fig.~\ref{fig:ht} (Left); The characteristic strain of GW signal by $h_c = 2f|h(f)|$ (Right) . The parameters are the same as in Fig.~\ref{fig:ht}.}
\label{fig:hf}
\end{figure*}

\begin{figure*}[htbp!]
    \centering
    \includegraphics[width=\linewidth,trim=1.0 1.0 1.0 1.0,clip]{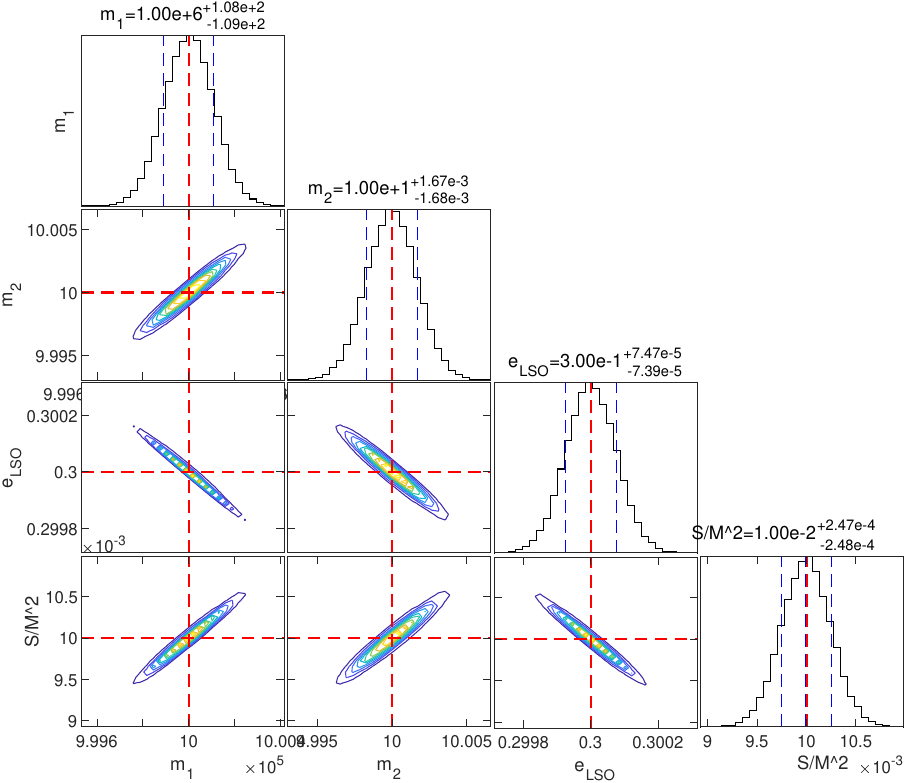}
    \caption{The corner plot from FIM exploration with fiducial/injected values $m_1=10^{6} M_{\odot}$ (central mass), $m_2=10 M_{\odot}$ (orbiting mass), $S/M^{2}=0.01$ (dimensionless spin of the central BH), $e_\mathrm{LSO}=0.3$. We have assumed an observation of $t_c = 3.14\times10^6 s$. redshift $z=0.01$, Median and $68 \%$ confidence interval are $m_1 =1_{-1.09\times10^{+2}}^{+1.08\times10^{+2}} \times 10^{6} M_{\odot}$, $m_2=10_{-1.70\times10^{-3}}^{+1.69\times10^{-3}} M_{\odot}$, $e_\mathrm{LSO} =0.3_{-7.39\times10^{-5}}^{+7.49\times{10^{-5}}}$, and $S/M^{2}=0.01_{-2.49\times10^{-4}}^{+2.47\times10^{-4}}$.}
    \label{Fig:MC_FM}
\end{figure*}

\subsection{GW Waveform}
Upon the execution of the \texttt{main.m} file, the \textbf{Common\_CF} module is activated, which facilitates the automatic retrieval of binary system parameters from the `input.txt' file (see Table~\ref{tab:parameters}). 

Following the loading of the waveform module, the \texttt{evalution( )} function is invoked, enabling a comprehensive analysis of the variations in the evolutionary orbital parameters of the binary system. Figure~\ref{Fig:evolution} shows the variation of the orbital parameters of the system: frequency, and eccentricity over time. One can observe that with the accumulation of duration time, the orbital frequency progressively increases. Conversely, due to the influence of GW radiation emitted by the system, the orbital eccentricity diminishes over time.

By executing the \texttt{waveform( )} function and selecting the response function in the \textbf{Detector.m} model, we can obtain the GW in time-domain (As shown in Figure~\ref{fig:ht}). Subsequently, applying a Fourier transform (\texttt{Fourier\_tran( )}) allows us to derive the frequency-domain waveform and the characteristic strain of the GW (As shown in Figure~\ref{fig:hf}).

\subsection{Fisher Information Matrix}
Before the execution of the \textbf{Fisher\_matrix.m}, it is imperative to first calculate the SNR of the GW signal. The FIM can be used with greater accuracy only when the SNR is sufficiently high. By integrating the characteristic frequency spectrum of the GW, we can obtain their SNR.

By executing the \texttt{diff\_param( )} function, we can obtain the partial derivatives of the GW signal for various parameters. By utilizing the \texttt{inner\_prod( )} function for the convolution between signals and applying Equation~\eqref{eq:gamma_vuale} to derive the $\Gamma$, we can then invoke the built-in MATLAB function \texttt{inv( )} to acquire the covariance matrix for the \mbox{different parameters.}

In Figure~\ref{Fig:MC_FM}, we plot the results of the posterior distributions for the central BH $m_1$, the small mass $m_2$, the eccentricity corresponding to the last stable orbit $e_{\mathrm{LSO}}$, and the spin of BH $\frac{S}{M^2}$. The one-dimensional marginal posterior distributions demonstrate that the FIM yields Gaussian profiles centered on the true parameters with standard deviations, as theoretically expected; the red dashed line represents the true value, while the thin blue dashed lines indicate the $1\sigma$ confidence interval. The two-dimensional joint distributions reveal parameter correlations through confidence regions represented by color-coded contours. Quantitative analysis of these bivariate distributions indicates strong correlations among all four parameters ($m_1$, $m_2$, $\frac{S}{M^2}$, and $e_{\mathrm{LSO}})$.

\begin{figure*}[htbp!]
    \centering
    \includegraphics[width=\linewidth]{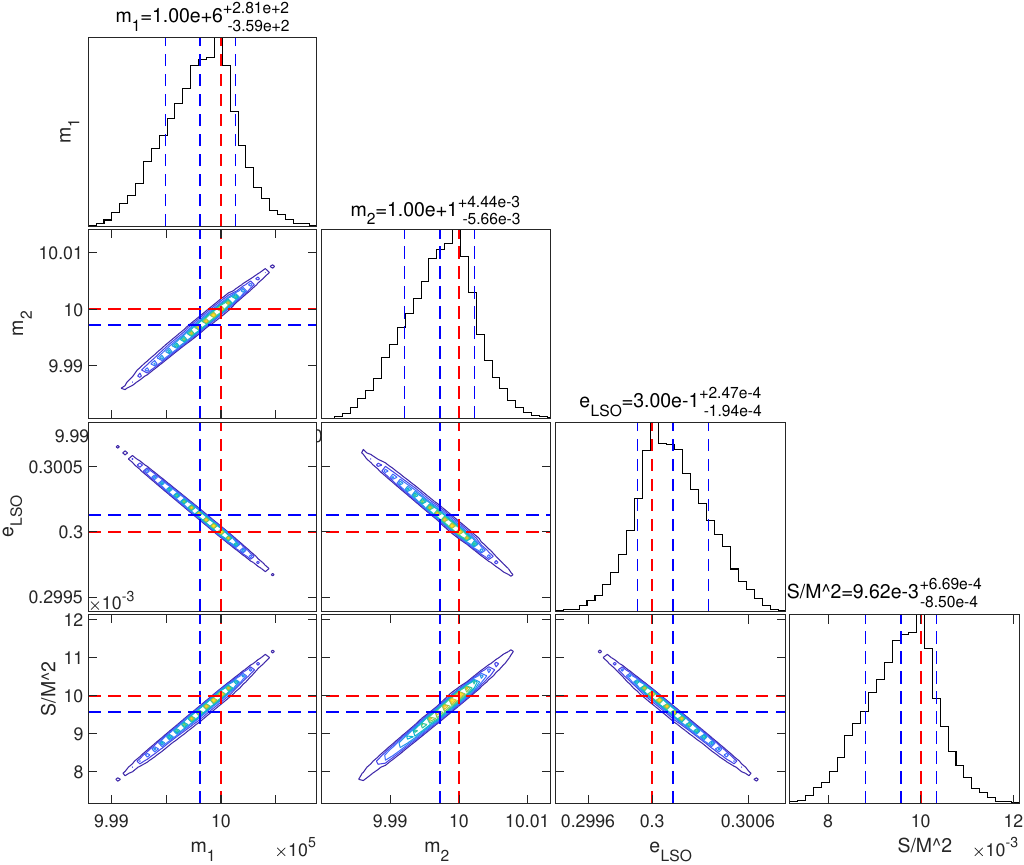}
    \caption{The corner plot from MCMC exploration with binary system whose parameters are the same as Fig.~\ref{Fig:MC_FM}, Median and $68 \%$ confidence interval are $m_1 =1_{-3.59\times10^{+2}}^{+2.81\times10^{+2}} \times 10^{6} M_{\odot}$, $m_2=10_{-5.65\times10^{-3}}^{+4.44\times10^{-3}} M_{\odot}$, $e_\mathrm{LSO} =0.3_{-1.94\times10^{-4}}^{+2.47\times{10^{-4}}}$ and $S/M^{2}=9.62 _{-8.50\times10^{-3}}^{+6.69\times10^{-4}} \times10^{-3}$.}
    \label{Fig:MC_MC}
\end{figure*}

\subsection{MCMC}
For parameters requiring Bayesian posterior estimation, we first define their prior probability distributions. $m_1\in\left[1.0 \times 10^5, 1.0 \times 10^7\right] \, M_\odot$, $m_2 \in \left[1.0, 1.0 \times 10^2\right] \, M_\odot$, $\frac{S}{M^2} \in \left[0.001, 0.1\right]$, $e_{\mathrm{LSO}} \in \left[0.2, 0.4\right]$, and the true values of parameters are the same as in Table~\ref{tab:parameters}. The posterior probabilities are then computed using the likelihood function \texttt{lpost( )}. We initialize multiple parallel Markov chains (typically set to $2k+1$ chains, where $k$ represents the dimensionality of parameters needing estimation). The sampling process is implemented through the \texttt{MCMC\_run( )} function following the M-H algorithm (Equation~\eqref{eq:M-H}). Convergence is monitored by calculating the vector $\hat{R}$ via the \mbox{\texttt{converge( )}} function. If $\hat{R} > 1.05$ (indicating non-convergence), we increase the sampling iterations and resume the process using \texttt{MCMC\_contin( )} until all $\hat{R} < 1.05$ criteria are satisfied. 

In Figure~\ref{Fig:MC_MC}, we present the posterior distributions for various parameters (\(m_1\), \(m_2\), \(e_\mathrm{LSO}\), and \(\frac{S}{M^2}\)). The blue dashed line indicates the median value, the thin blue lines denote the credible intervals \(68\%\) derived from the posterior distribution, and the red dashed line represents the true parameter values. Our analysis reveals that the results estimated via MCMC show good agreement with those obtained from the FIM in Figure~\ref{Fig:MC_FM}. Although the MCMC-derived posteriors exhibit marginally broader confidence intervals, they reflect a more complete exploration of the parameter space.

To validate the accuracy and computational efficiency of \texttt{MatBYIB}, we compared the $68\%$ credible intervals of parameter estimates obtained by FIM, \texttt{mcmcstat}, and \texttt{MatBYIB}. As for \texttt{MatBYIB}, we implemented parallel computing using the explicit \texttt{parfor} loops in MATLAB's Parallel Computing Toolbox (version R2020a). All computations were executed on an Intel\textregistered\ Core\texttrademark\ i5-10400 processor (6 physical cores, and 12 logical threads with a base frequency of 2.90 GHz), with 10 worker threads allocated to optimize computational efficiency while ensuring system stability. A total of 10 independent chains were run, with each generating 10,000 samples. The computation achieved convergence ($\hat{R}<1.05$) after 80,189.70 s ($\sim$22.3 h). Under identical conditions, \texttt{mcmcstat} completed sampling in 5.3 h.

As shown in Table~\ref{tab:convergence}, the FIM achieves the highest precision with
$\Delta m_1 \in [-1.09, 1.08] \times 10^{2}$,
$\Delta m_2 \in [-1.68, 1.67] \times 10^{-3}$,
$\Delta\frac{S}{M^2} \in [-7.39, 7.47] \times 10^{-5}$, and
$\Delta e_{\mathrm{LSO}} \in [-2.48, 2.47] \times 10^{-4}$.
The \texttt{mcmcstat} implementation yields
$\Delta m_1 \in [-2.60, 3.65] \times 10^{2}$,
$\Delta m_2 \in [-3.99, 5.70] \times 10^{-3}$,
$\Delta\frac{S}{M^2} \in [-2.52, 1.77] \times 10^{-4}$, and
$\Delta e_{\mathrm{LSO}} \in [-5.93, 8.49] \times 10^{-4}$.
For \texttt{MatBYIB}, convergence was achieved at 10,000 iterations ($\hat{R} \approx 1.046$ for all parameters), producing refined uncertainties:
$\Delta m_1 \in [-3.59, 2.81] \times 10^{2}$,
$\Delta m_2 \in [-5.66, 4.44] \times 10^{-3}$,
$\Delta\frac{S}{M^2} \in [-1.94, 2.47] \times 10^{-4}$, and
$\Delta e_{\mathrm{LSO}} \in [-8.50, 6.69] \times 10^{-4}$.
The results show general agreement between all three methods, with FIM providing theoretically optimal errors. \texttt{MatBYIB} demonstrates slightly superior precision compared to \texttt{mcmcstat}, albeit with higher computational cost.

\begin{table*}[htbp]
\caption{$68\%$ central intervals and estimated Potential Scale Reduction Factors for four scalar sum
maries of the multivariate normal distribution simulated using a Metropolis algorithm. Displayed are inferences from the second halves of nine parallel sequences, stopping
after 2000, 5000, and 10000 iterations. The intervals for ($\infty$) are taken from the known normal distributions for these summaries in the target distribution.\label{tab:convergence}}
\begin{tabular}{ccccc}
\toprule
\textbf{Iteration} & \textbf{$m_1,R$} & \textbf{$m_2,R$} & \textbf{$e_\mathrm{LSO},R$} & \textbf{$S/M^2,R$} \\
\midrule
FIM & $[-1.09,1.08]\times10^2$&$ [-1.68,1.67]\times10^{-3} $&$ [-7.39,7.47]\times10^{-5} $&$ [-2.48,2.47]\times10^{-4}$\\
\texttt{mcmcstat}&$ [-2.60,3.65]\times10^2 $&$ [-3.99,5.70]\times10^{-3}$&$ [-2.52,1.77]\times10^{-4}$&$ [-5.93,8.49]\times10^{-4}$\\
$2000 $&$ [-3.48,3.89]\times10^2, 1.828 $&$ [-5.48,6.16]\times10^{-3}, 1.832 $&$ [-2.68,2.39]\times10^{-4},1.824 $&$ [-8.15,9.13]\times10^{-4},1.827$ \\
$ 5000 $&$ [-3.52,3.13]\times10^2, 1.146 $&$ [-5.51,5.45]\times10^{-3}, 1.146 $&$ [-2.13,2.44]\times10^{-4},1.146 $&$ [-8.32,8.53]\times10^{-4},1.146$ \\
$ 10000 $&$ [-3.59,2.81]\times10^2, 1.046 $&$ [-5.66,4.44]\times10^{-3}, 1.045 $&$ [-1.94,2.47]\times10^{-4},1.046 $&$ [-8.50,6.69]\times10^{-4},1.046 $\\
\bottomrule
\end{tabular}
\end{table*}

\section{Conclusions}
We have developed a simple and user-friendly MATLAB-based code package for the Bayesian analysis of GW parameters for binaries with arbitrary eccentricities, \texttt{MatBYIB}. This package is based on the AK waveform, which employs the post-Newtonian approximation to simulate the orbital parameter evolution and utilizes the Peters--Mathews formalism~\cite{Peters1963} to obtain the GW quadrupole moment. Then, FIM and MCMC are both employed to reconstruct the posterior distributions of GW parameters. \texttt{MatBYIB} incorporates parallelization, enabling routine numerical simulations of PE on modest hardware (such as a desktop computers) without compromising numerical convergence. We demonstrated its potential applications through examples of GW parameter estimation for EMRI binaries in elliptical orbits. Comparisons between results obtained from the FIM and those from MCMC show good agreement.

Given the excellent scalability of the AK waveform, future developments of \texttt{MatBYIB} could include additional effects, such as dynamical friction from dark matter~\cite{Eda2013,Li_2022}. To enhance the accuracy of the waveform, higher-order post-Newtonian terms could be incorporated, or alternative waveform models such as \texttt{FastEMRI}~\cite{Katz2021} could be adopted.


\end{document}